# Algorithmic Fairness in Education
René F. Kizilcec and Hansol Lee (Cornell University)

**Introduction**

Educational technologies increasingly use data and predictive models to provide support and analytic insights to students, instructors, and administrators (Baker & Inventado, 2014; Luckin & Cukurova, 2019). Adaptive systems like cognitive tutors help students achieve content mastery by providing them with different study materials based on predictions about what content they have already mastered (Pane et al., 2010). Automated scoring systems provide immediate feedback on open-ended assessments based on predictions of what scores and comments a human grader would give (Yan et al., 2020). Student support systems identify struggling students to automatically offer them assistance or flag them to instructors or administrators based on predictions about which students are likely to disengage from the learning platform, get a low score on an upcoming assessment, or experience affective states of confusion, boredom, and frustration (Hutt, Grafsgaard, et al., 2019; Prenkaj et al., 2020). Some educational technologies use data-driven predictions to directly change the learning experience, such as skipping over a module that a student is predicted to have mastered already. This can occur with or without explicit notification to the student, rendering the "intelligence" of a system either overt or hidden. Other educational technologies present model predictions to students, instructors, or administrators to support their process of interpretation and decision making. The presentation format of such predictions varies substantially with the learning context, target audience, and desired response; it can be in the form of a dedicated dashboard for instructors to track students or for students to monitor their own progress, indicators embedded in learning activities for immediate feedback, or subtle changes in the digital learning environment that affect student attention and behavior. The influence of artificial intelligence in education is growing in K-12, higher, and continuing education with the increasing adoption of algorithmic systems that employ predictive models developed using big data in education.

The increasing use of algorithmic systems in education raises questions about its impact on students, instructors, institutions, and society as a whole. How much and under what circumstances do such technologies benefit these stakeholders? What characteristics of algorithmic systems in education are associated with greater benefits? And what counts as a beneficial impact on these stakeholders? These questions encourage a critical analysis of AI in education that also attends to its perhaps unintended and unforeseen negative consequences. Those have been a subject of significant public and academic discourse in other domains including credit decisions, employment screening, insurance eligibility, marketing, delivery of government services, criminal justice sentencing and probation decisions (MacCarthy, 2019; O'Neil, 2016). The use of data-driven decision systems in these domains has raised concerns about fairness, bias, and discrimination against members of protected classes in the United States (women, seniors, racial, ethnic, religious, and national minorities, people with disabilities, genetic vulnerabilities, and pre-existing medical conditions). It is time to widen the scope of this research effort to examine potential issues of fairness arising from the use of algorithmic systems in educational contexts. This chapter is an introduction to algorithmic fairness in education with a focus on how discrimination emerges in algorithmics systems and how it can be mitigated by considering three major steps in the process: measurement (data input), model learning (algorithm), and action (presentation or use of output). Building on extensive research on algorithmic fairness in other domains, we examine common measures of algorithmic fairness, most of which focus on the model learning, and how a better understanding of threats to algorithmic fairness can advance the responsible use of artificial intelligence in education. In another chapter in this book, Holstein and Doroudi (2021) holistically examine the socio-technical system surrounding artificial intelligence in education to explore how it risks amplifying inequalities in education.

**Fairness in Education**

Considerations of fairness are deeply rooted in the field of education and focused on concerns of bias and discrimination. Long before the adoption of digital learning environments in schools and homes, education scholars have studied inequalities and inequities in educational opportunities and outcomes, such as school segregation and achievement gaps. In fact, this work foreshadowed more recent formal definitions of algorithmic fairness in machine learning (Hutchinson & Mitchell, 2018). In 1954, the U.S. Supreme Court stated in *Brown v. Board of Education*, "it is doubtful that any child may reasonably be expected to succeed in life if he is denied the opportunity of an education." The legal recognition of the value that an education provides and the state's role in its provision put an end to state-sanctioned school segregation in the U.S. After all, if state-supplied education is of such critical value, according to *Brown v. Board of Education,* it "must be available to all on equal terms." This ruling shaped decades of research and public discourse on equal opportunities to educational access (Reardon & Owens, 2014), and it can be understood as a requirement for fairness in educational access. At least since the Coleman report of 1966, academic achievement gaps have become a focus of educational reform efforts (Coleman, 2019). Colman and many others have argued that a combination of home, community, and in-school factors give rise to systematic differences in educational performance between groups of students based on their socioeconomic status, race-ethnicity, and gender (Kao & Thompson, 2003). The presence of achievement gaps can be understood as a shortcoming of fairness in educational outcomes, especially if it is the result of discriminatory behavior: it is unfair if students from low-income families score lower test scores for lack of access to study resources available to high-income families, but it is especially unfair if they score lower because their teacher -- or an algorithmic scoring system -- is biased against them. Given the long tradition of scholarship on inequities in education, the term algorithmic fairness in the academic community almost exclusively refers to bias and discrimination involving algorithmic systems.

Any notion of fairness is inherently based in social comparison. The impact of an educational policy or technology can be assessed for groups of individuals, and how the impact compares in magnitude between them has fairness implications. Figure 1 shows three stylized representations of how an innovation, such as the use of an intelligent tutoring system for mathematics, might affect educational outcomes for groups and individuals under the assumption that the innovation is beneficial in general and that there are pre-existing gaps in outcomes. Each panel shows how an innovation can affect members of an advantaged and a disadvantaged group on average. Although the average outcome improves in both groups, the relative level of improvement in the advantaged group here determines whether the innovation reduces the pre-existing gap (closing gap), expands it (widening gap), or leaves it unchanged (constant gap). Most studies of the impact of educational technology on student outcomes find evidence consistent with the widening gap scenario (Attewell & Battle, 1999; Boser, 2013; Warschauer et al., 2004; Wenglinsky, 1998), though there are some exceptions that offer evidence of constant or even closing gaps (Rene F. Kizilcec et al., 2021; Roschelle et al., 2016; Theobald et al., 2020). Most observers would agree that a widening gap is less fair than a constant gap, and some may find a constant gap to be less fair than a closing gap even though the overall improvement in outcomes is lower. All three scenarios are certainly fairer than one where the disadvantaged group does not benefit at all or even sees decrements in average outcomes. Fairness should therefore be measured on a continuous and not binary scale, and there are multiple ways to measure fairness, as we will discuss in more detail. Note that the points presented in Figure 1 represent group averages, which can mask important distributional group differences (e.g., spread and skew of outcomes within groups) that can raise additional fairness concerns and warrant closer inspection in practice.

Equality and equity are two common notions of fairness in education that are also represented in Figure 1. An innovation achieves equality in its impact if the groups benefit the same amount no matter what their pre-existing outcomes are (constant gap); but to achieve equity, its impact has to be more beneficial to the group with lower outcomes to close pre-existing gaps (closing gap). Most observers would ascribe different levels of fairness to these two outcomes. In one case, groups of students who are different in terms of their outcomes (and perhaps in other ways) receive the same benefit and remain different; in the other case, different groups of students receive different benefits to render them more alike. But what about students who are similar to begin with? Dwork and colleagues (2012) propose a definition of individual fairness whereby similar individuals are to be treated similarly. Albeit a simple and intuitive approach to measure fairness, it raises a different challenge in adequately measuring the similarity of individuals: which individual characteristics are considered of those that are available and how they are combined can interfere with the assessment of individual fairness. Most definitions of fairness used in practice are therefore at the group level and compare the average individual across different groups; examples in education include the assessment of enrollment and graduation rates and achievement gaps.

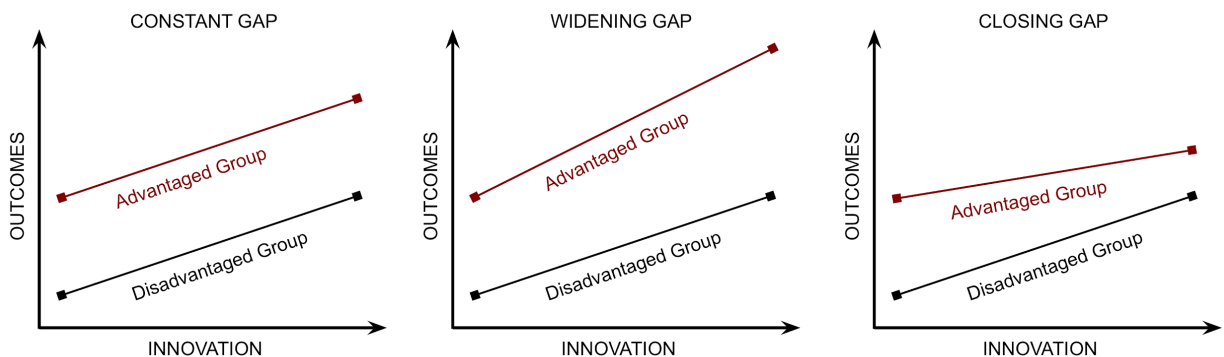

**Figure 1.** Three stylized representations of how a generally beneficial innovation can influence outcomes for members of an advantaged and a disadvantaged group.

Scholars have grappled with issues of algorithmic fairness in high-stakes settings like healthcare and criminal justice, where machine learning models have become widely adopted in practice (Corbett-Davies

& Goel, 2018; Lum & Isaac, 2016; Wiens et al., 2019). Although considerations of fairness in education are not novel, the consideration of *algorithmic* fairness is more recent and motivated by the growing number of students who are affected by algorithmic systems in educational technologies today (Hutchinson & Mitchell, 2018). This chapter provides an introduction to algorithmic fairness in education by examining the components of an algorithmic system and where considerations of fairness enter in the process of developing and deploying these systems. Our focus is on algorithmic fairness defined by the absence of bias and discrimination in a system, rather than the presence of due process. An algorithmic system such as college admission by random number generator is unfair for using an arbitrary and unaccountable process, but it is still unbiased and non-discriminatory. We review different ways of defining and assessing fairness, and initial evidence on algorithmic fairness in education. We conclude by offering recommendations for policymakers and developers of educational technology to promote fairness in educational technology.

**Algorithmic Fairness and Antidiscrimination**

A fair algorithm does not discriminate against individuals based on their membership in protected groups. Then what does it mean for an algorithm to discriminate and what are protected groups? The latter question can have a legal answer, which in countries like the US or UK includes groups defined based on age, sex, color, race, religion, nationality, citizenship, veteran status, genetic information, and physical or mental ability (Government Equalities Office, 2013). Yet the list of protected groups can be extended or modified to fit the specific application of an algorithmic system. Ocumpaugh and colleagues (2014) tested the accuracy of a student affect detector on students in urban, suburban, and rural schools. Their interest in algorithmic fairness focused on protected groups defined based on urbanicity of students' location, because their affect detector was developed with data from predominantly urban students and would eventually get adopted in more rural schools. Doroudi and Brunskill (2019) tested how an intelligent tutoring system affects learning outcomes for fast and slow learners. Their interest in algorithmic fairness focused on protected groups defined based on students' learning speed to evaluate if the self-paced system discriminates against students from either group. Although definitions of protected groups vary across applications and contexts, it is necessary to specify a priori which protected groups' algorithmic fairness is to be determined.

To explain what it means for a system to discriminate against individuals, we consider traditional forms of discrimination involving human agents, which may be intentional or unintentional in nature. American antidiscrimination laws distinguish between disparate treatment and disparate impact to draw a distinction based on an actor's intent. For example, in the context of college admissions, direct consideration of applicants' race in admissions decisions (outside of holistic review) is a form of disparate treatment, because the rule that is being applied is not neutral with respect to a protected attribute. If applicants' race is omitted from consideration in the process and yet disproportionately fewer applicants from some racial group are admitted, it is a form of disparate impact but not treatment, which does not require proof that it was intentional. Now, consider a college admissions algorithm that ranks applicants according to their predicted academic performance based on historical data from applicants and their subsequent college achievement. If this algorithmic system ranks applicants from one racial group disproportionately lower than others, it may be said that it discriminates against individuals in that protected group. Barocas and Selbst (2016) argue that this is almost always due to preexisting patterns of bias in historical data or unintentional emergent properties of the system's use, rather than conscious choices by its programmers. In the absence of demonstrable intent, the system's discrimination is a case of disparate impact, not treatment. Including protected attributes as inputs into the algorithm does not typically change its discriminatory effects, and without a case for intentional discrimination, it does not amount to disparate treatment. Most unfair algorithmic systems thus discriminate against individuals in unintentional ways that create disparate impact. We acknowledge the importance of addressing systemic patterns of injustice that contribute to unfair algorithms and the hazards of treating the status quo as fixed. At the same time, computational research on algorithmic fairness in education can play a valuable role in affecting social change (Abebe et al., 2020).

**How Discrimination Emerges in Algorithmic Systems**

In the absence of discriminatory intent, how do algorithmic systems produce disparate impact among protected groups? To answer this question, we deconstruct how a generic algorithmic system is developed and used, and identify how issues of fairness can arise in the process even without deliberate intent to discriminate. A typical data-driven algorithmic system makes predictions about future or previously unseen cases based on what it "learns" from historical data. The development and use of the system can be deconstructed into a sequence of three steps visualized in Figure 2: measurement, model learning, and action. Measurement is the process of collecting data about an environment. Model learning is the process of using the collected data (i.e. training data) to develop a representation of the environment as a set of correlations. Action is the process of using predictions of the learned model for new cases for judgement and decision making. It can be a system-based action or a human action by a number of different stakeholders, including students, instructors, teaching staff, and administrators.

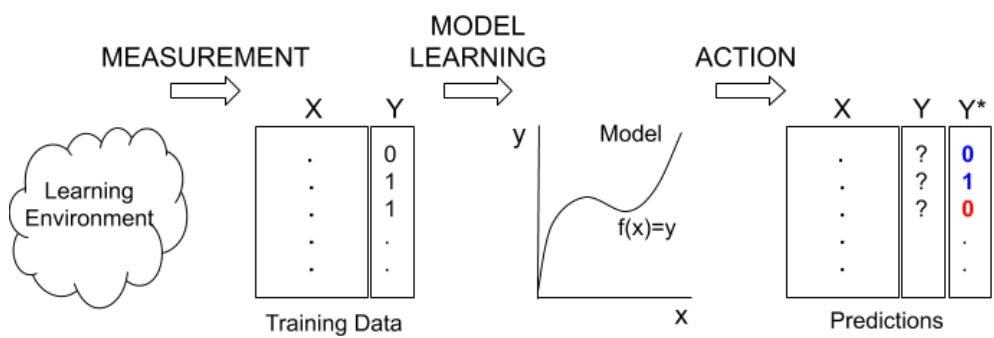

**Figure 2.** Development and use of a generic algorithmic system: Data is collected from a learning environment of interest, with each datapoint represented by some fixed set of attributes $X$ with an outcome of interest $Y$ (Measurement); the collected data is used to learn a model $f$ that represents the relationship between $X$ and $Y$ (Model Learning); the learned model is used to predict outcomes $(Y^*)$ for new data with unknown true outcome $Y$ to inform system or stakeholder actions (Action).

We return to the example of an algorithmic system designed to support the college admissions process by predicting students' college success from their application data (Friedler et al., 2016; Kleinberg et al., 2018). In the measurement step, a training dataset is assembled by quantifying college success (e.g., cumulative GPA, degree completion) as the target variable $Y$ for each student $s_i$ with a set of individual features $\{x_1, ..., x_n\} \in X$ from the application information. Student features may include high school grades, standardized test scores, the number of extracurricular activities and leadership positions, and linguistic features extracted from the essay. Once the training dataset is assembled, the model learning step can begin. There is a wide variety of prediction algorithms to choose from. For instance, a (penalized) linear regression algorithm might be used to learn a model for a continuous outcome like cumulative GPA, while a random forest algorithm might be used to learn a model for a binary outcome like on-time graduation. The learned model is then used to generate predictions for new applicants whose eventual college success is yet unknown. These predictions can be used in the action step to inform judgement and decision-making about applicants in a number of ways. Admissions officers may still review all applicants manually, but take the model predictions into account as an additional input into their decisions. Alternatively, the algorithmic system may provide admissions officers the option to further review only applicants whose predicted college success rate is above a certain threshold. In theory, following careful evaluation of its feasibility and impact, the entire admissions process could be automated based on the model predictions. Using this example of a college success prediction system, we examine how unintended discrimination can arise in each of the three steps visualized in Figure 2.

*Measurement*

Measurement may appear like the simplest step in the process, but it presents some of the greatest challenges for developing fair algorithmic systems. The first task is to define the prediction problem and measure the target variable. The number of recent articles emphasizing the importance of defining appropriate outcomes in educational settings and the dangers of optimizing for inappropriate ones is evidence that this task calls for careful consideration of potential options and their consequences (DeBoer et al., 2014; Duckworth & Yeager, 2015; Gašević et al., 2015; Schwartz & Arena, 2013). In the context of college admissions, selecting a target measure of college success is not only a task that requires subjective judgements of what constitutes success (academics, extracurriculars, community service, etc.), it also depends on the characteristics and objectives of the admitting college (its graduation rates, goals for student diversity, etc.). Consider college GPA as a target measure of college success, which has been used in prior research (e.g., Kleinberg et al., 2018). This measure arguably captures a limited notion of college success, as students can be successful in other ways than achieving high grades in their classes. Most target measures, including college GPA, inherit prejudices of prior decision-makers and encode widespread biases that exist in society. For instance, persistent demographic gaps in the GPA of college graduates may not reflect true differences in college success (de Brey et al., 2019): structural barriers like access to learning resources, psychological factors like stereotype threat, and (implicit) bias among graders can all reduce GPA for historically disadvantaged students. A learning algorithm will discover the correlation between student demographics and college GPA in the training data, even though it is a reflection of historic patterns of prejudice, discrimination, or data integrity issues. The subjectivity of the choice of the target variable and the possibility that it encodes bias can adversely affect the fairness of an algorithmic system. The target variable is particularly influential in this regard because it alone encodes what the algorithm optimizes for, but the choice and measurement of features presents its own set of challenges (Nabi & Shpitser, 2018).

The selection of features to measure alongside the target measure involves a reduction of a rich state of the world into a fixed and relatively narrow set of values. Due to this distillation, the collected data won't capture the full complexity of individual differences and contextual factors. This raises concerns for algorithmic fairness if historically disadvantaged and vulnerable members of society are differentially affected in the process. For instance, the inclusion of Advanced Placement (AP) grades as a feature in the training data for an algorithmic admissions system can result in bias against students who do not have access to AP classes in their high school, which disproportionately affects students in low-income school districts. Moreover, historical patterns of bias can be embedded in measured features in the same way as the target variable. Standardized test scores such as ACT or SAT scores are commonly used in the admissions process as an indicator of academic potential, even though they are highly correlated with students' socioeconomic status and could therefore disadvantage minority and low-income students (Sackett et al., 2009). The predictive power of features can also vary across groups. For example, features extracted from the college admissions essay may vary in how well they forecast college success depending on students' native language and home country, as evidenced by research demonstrating cultural variation in the predictive power of motivational factors on student achievement (Li et al., 2021).

Beyond the problem definition, the process of measurement almost always requires sampling data from a population. Whether sampling is done intentionally or unintentionally, it warrants careful consideration because it raises questions of representativeness and generalizability. These terms are often vaguely defined and used inconsistently across fields such as politics, statistics, and machine learning (Chasalow & Levy, 2021). In general, the closer the training data is to the test data for which the algorithm is going to make predictions, the more accurate its predictions are going to be; however, this is not always the case (Yudelson et al., 2014). Ocumpaugh and colleagues (2014) demonstrated this in the case of a student affect detector that exhibited higher prediction accuracy for students from rural, suburban, and urban regions if it was trained on a sample drawn from the same locale. Likewise, Gardner and colleagues (2019) showed in the context of student dropout prediction that training datasets that skew males provide lower prediction accuracy for females. The underrepresentation of some groups in the training data due to the sampling strategy can thus present a threat to algorithmic fairness by disadvantaging members of historically underrepresented groups (Chawla et al., 2002).

*Model learning*

The model learning step typically begins with preprocessing the collected training data. Preprocessing can take many different forms and commonly includes the removal of duplicate data, correction of inconsistencies, removal of outliers, and handling of missing data. According to a recent review paper on predictive student modeling, very few studies provide a detailed account of the preprocessing procedures that were applied (Cui et al., 2019). Friedler and colleagues (2019) show that different choices made during preprocessing can lead to notable differences in the performance of algorithmic systems and should therefore be well-documented and held constant when comparing model learning strategies.

Model learning is the process of approximating the relationship between the features and the target measure based on the preprocessed training data. The resulting model is therefore subject to potential biases embedded in the data. Some biases can be mitigated in the measurement step itself, but others will enter into the model learning process and without an intervention, the resulting model is likely to reflect these biases. To illustrate this point, consider that without intervention, a natural language model trained on textual data collected from a large real-world corpus like Google News will learn gender stereotypes such as "male" is to "computer programmer" as "female" is to "homemaker" (Bolukbasi et al., 2016). This shows that even if the training dataset is remarkably large and the embedded biases are relatively subtle, the learned model is still likely to mirror the bias in its predictions if left unchecked. Studies in education that found reduced prediction accuracy for students who are underrepresented in the training data could for instance use case weights to raise the relative influence of underrepresented students in the model learning process (Gardner et al., 2019; Ocumpaugh et al., 2014).

Two major choices in the model learning step are the type of model and evaluation strategy. In practice, most software packages render it quick and easy to try different types of models and compare their performance against each other: for example, *scikit-learn* in Python (Pedregosa et al., 2011) and *caret* in R (Kuhn & Others, 2008). Early research on how the choice of a type of model affects algorithmic fairness does not suggest that some model types are generally fairer than others; instead, the level of algorithmic fairness varies across datasets, and even within datasets for different random splits into training and testing data (Friedler et al., 2019; Gardner et al., 2019). This finding emphasizes the role of the model evaluation strategy and metrics in assessing algorithmic fairness. A common evaluation strategy is to compute model performance as the overall prediction accuracy on a held-out testing dataset. However, the high overall accuracy can hide the fact that the model has much lower accuracy for individuals who are underrepresented in the data. Gardner and colleagues (2019) advocate for a "slicing analysis" to consider model accuracy for subgroups explicitly and quantify the gap in accuracy. Others have proposed modifications to learning algorithms to additionally optimize a fairness constraint (Calders & Verwer, 2010; Kamishima et al., 2012; Zafar et al., 2017; Zemel et al., 2013). How analytic choices in model learning raise issues of algorithmic fairness is a relatively novel and active area of research, and much remains to be learned (for a recent survey, see d'Alessandro et al., 2017).

*Action*

The final step in the process is taking action using the learned model to make predictions for new cases and guide human decision-makers (students, instructors, administrators, etc.) or have the algorithmic system act upon the predictions directly. In the case of an algorithmic system for college admissions, predictions for college success will be made for new applicants who were not in the training dataset and for whom the true outcome such as college GPA is unknown. The resulting predictions could be used in various ways, from merely complementing the standard application materials that admissions offers review, to entirely automating admissions decisions. The predictions are however only as valid as the underlying training data and learned model. If biases in the underlying data result in biased predictions, for instance underpredicting college success for minority and low-income students, it can affect admissions decisions and contribute to disparate impact. It is therefore recommended to monitor the accuracy of model predictions over time and set expectations that the model will likely require tuning when fairness issues are discovered. Failing to test an algorithmic system for potential fairness issues, known as auditing the algorithmic system, can result in disparate impact (Saleiro et al., 2018). Yet the growing complexity of algorithmic systems is making it more difficult to test them.

As algorithmic systems have grown in complexity, it has become more difficult for human decision-makers who use these systems to understand why and how a model is making certain predictions. The rise of "black box" systems has raised concerns about the trustworthiness of model predictions, the extent to which predictive models should also be explanatory in nature, and what happens when predictions drive automated decisions that may have discriminatory effects (Hosanagar, 2020). These concerns have motivated research on *interpretable machine learning* which aims to design models that are transparent and understandable (Conati et al., 2018; Doshi-Velez & Kim, 2017). If predictive models provided explanations for their predictions, such as why a given student applicant is predicted to have low college success, then decision-makers could use these explanations to qualitatively assess various criteria including fairness. Concerns about model transparency and explainability have come up in the knowledge tracing literature, which has been dominated by simple and interpretable Bayesian models (Corbett & Anderson, 1995) when a new deep learning model was proposed with substantially more parameters and lower interpretability (Piech et al., 2015). How to communicate the prediction outcomes to the decision-maker or user of the system is a human-computer interaction question. Deciding just how much transparency to provide about an algorithmic system is a critical consideration that can determine how much users will trust the system and its predictions (René F. Kizilcec, 2016). An erosion of trust in an algorithmic system can lead decision-makers to disregard its predictions, which can result in discriminatory action that reinforces prevailing stereotypes when decisions are also subject to confirmation bias (Nickerson, 1998). For example, an admissions officer who does not have trust in the algorithmic system may question its predictions when they violate her expectations (stereotypes) but not otherwise.

Another potential issue that can arise even if the model is thoroughly tested and highly accurate, is the misinterpretation of a fundamentally correlational prediction as a causal one. Most algorithmic systems are developed based on historical data for which they learn to represent empirical relationships between features and the target measure. Model predictions therefore indicate correlational but not causal quantities in most cases; and yet, causal interpretations of predictive models are abundant. SAT scores are predictive of college success in terms of GPA, but it does not mean that if SAT scores were raised for some students by signing them up for a test preparation course, that it would cause them to earn a college GPA according to the prediction. Instead, numerous individual and contextual factors about a student correlate with both SAT score and college GPA to strengthen the correlation between these two measures. Using said predictions to support the college admissions process may therefore result in disparate impact, by more frequently denying admission to low-income students who tend to have both lower SAT scores and college GPA (Sackett et al., 2009). This provides another potential source of unfairness in algorithmic systems, especially if it establishes a negative feedback loop by biasing future training data used to update the model of the algorithmic system (O'Neil, 2016).

For some prediction problems, it may not be possible to achieve a high level of prediction accuracy even with access to a rich dataset and state-of-the-art machine learning algorithms. Salganik and colleagues (2020) demonstrated this in a competition involving hundreds of researchers to predict major life outcomes including educational achievement based on years of detailed longitudinal survey data. The best performing model was only slightly better than a simple baseline and explained only 20% of the variation in high school GPA. Predicting who should receive what kind of educational intervention can be even harder because it requires causal inference. Kizilcec and colleagues (2020) used data from a massive field experiment in online higher education and state-of-the-art machine learning to target behavioral science interventions to individual students, but even the best models produced no better student outcomes than simply assigning everyone the same intervention or a random one. These studies highlight practical limits of data-driven models for prediction and potential risks of stakeholders placing too much confidence in algorithmic systems.

**Measures of Algorithmic Fairness**

The previous section showed how issues of fairness can arise in *every* step of the process of developing and deploying an algorithmic system. This can result in discriminatory action and disparate impact in the absence of any malicious intent. In this section, we take a closer look at what unfairness means by reviewing a number of formal definitions of fairness that have been proposed in the literature to date and how they can be applied in the context of education. Specifically, we review statistical, similarity-based, and causal notions of fairness in relation to an example of an algorithmic system that predicts student dropout. Research on fairness in machine learning is rapidly evolving and this review is based on the most recent work in the field (for further reading, we recommend Barocas and colleagues' online book on fair machine learning and Verma and Rubin's detailed review of definitions; Barocas et al., 2019; Verma & Rubin, 2018). Measures of algorithmic fairness have focused on the predictions resulting from the model learning step in the process of developing an algorithmic system. This step lends itself to quantifying fairness because of the standardized approaches to evaluating models, such as a confusion matrix of correct and incorrect predictions. In contrast, to quantify bias and discrimination in the measurement and action steps, significantly more contextual knowledge is required to quantify dataset bias (e.g., what population it should be representative of), discriminatory problem definitions (e.g., predicting affective states that are subject to cultural variation), stigmatizing responses to model predictions (e.g., flagging at-risk students in ways that discourages performance), and biased representations of predicted outcomes (e.g., visualizations that make small predicted differences seem large).

To compare different definitions of fairness in the context of AI in education, we consider the case of student dropout prediction which has received substantial research attention in various learning environments. Let random variables $X$ represent a set of observed features for each student, $Y$ their true dropout outcomes, $D$ the algorithmic decisions (predictions of $Y$), and $G$ a protected attribute of each student. For standard dropout prediction, this is a binary classification problem where $Y = \{0, 1\}$, though it can be generalized to multivariate or regression problems. The algorithmic system uses a prediction model $f(X)$ that returns a probability distribution over the possible values of $D$ for each individual (i.e. $Pr(D|X)$). $D$ is determined using a deterministic threshold $t$, where $D = 1$ if $f(X) > t$, and 0 otherwise. We examine whether an algorithmic system that predicts the likelihood of students' dropout from a course is fair for male and female students (i.e. $G$ in this case).

*Statistical notions of fairness*

We begin by reviewing three statistical notions of fairness: independence, separation, and sufficiency. They are a foundation for understanding many statistical fairness criteria in the literature, which can be expressed as a derivation of one of these three definitions. **Independence** requires that an algorithm's decision be independent of group membership. Formally, in the case of binary classification, it requires that $Pr(D=1|G=g_i) = Pr(D=1|G=g_j)$ for all protected groups $g_i$, $g_j$ in G. Independence is satisfied if the same percentage of male and female students are classified as at-risk for dropping out ($D=1$). Figure 3 visualizes outcomes of a dropout prediction algorithm that satisfies independence. Independence represents the desirable long-term goal of equity for different groups of students (i.e. male and female students persist at equal rates) and using it as a measure of fairness may advance efforts to reduce historical inequalities (Räz, 2021) However, a limitation of independence as a notion of fairness in settings like dropout prediction is that it ignores students' true tendency to drop out. If female students were more likely to drop out in the first week of class than male students, for instance because they initially enroll in more courses to find the best fit, then it would seem reasonable to account for the actual gender difference in dropout in the predictions. To address this limitation of independence, which results from only considering the protected group $G$ and algorithmic decision $D$, the definition of fairness as separation additionally considers the true outcome $Y$.

**Separation** requires that an algorithm's decision be independent of group membership conditional on true outcomes. Formally, in the case of a binary classification, separation requires that $Pr(D=1|Y=1, G=g_i) = Pr(D=1|Y=1, G=g_j)$ and $Pr(D=1|Y=0, G=g_i) = Pr(D=1|Y=0, G=g_j)$ for all protected groups $g_i$, $g_j$ in G. Separation encodes the belief that a fair algorithm makes correct and incorrect predictions at similar rates for different groups. This is visualized in Figure 4 (left panel) which shows the same true positive rate (60%) and false positive rate (20%) for male and female students. If the dropout prediction algorithm has a higher false positive rate for female students than for male students, it falsely flags well-performing female students as at-risk more often than well-performing male students. Instructors may lower their expectations about students flagged as at-risk and act towards them in ways that reduce their academic performance -- a phenomenon known as the Pygmalion effect (Brookover et al., 1969). This describes one way that an algorithmic system with a higher false positive rate for female students can be unfair towards female students. Likewise, if the dropout system has a lower true positive rate for female students than male students, it will fail to identify struggling female students more often than similarly struggling male students. A targeted intervention to help students predicted to be at-risk may inadvertently help male students more than it helps female students. To achieve separation, an algorithmic system typically needs to set different decision thresholds $t$ for each protected group. However, this violates another notion of fairness such as individual fairness, according to which individuals should not be treated differently on the basis of protected attributes.

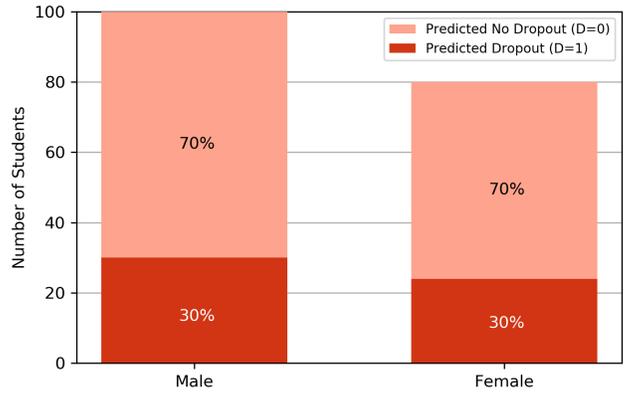

**Figure 3.** Illustration of dropout predictions that satisfy fairness as independence. Given a total of 100 male and 80 female students, independence is satisfied with $Pr(D=1|G=\text{male}) = Pr(D=1|G=\text{female}) = 30\%$ predicted to at-risk of dropping out.

Whereas separation requires that algorithmic decisions are independent of protected attributes conditional on true outcomes, **sufficiency** requires that true outcomes are independent of protected attributes conditional on algorithmic decisions. Formally, in the case of a binary classification, sufficiency can be expressed as $Pr(Y=1|D=1, G=g_i) = Pr(Y=1|D=1, G=g_j)$ for all protected groups $g_i$, $g_j$ in G. Sufficiency encodes the belief that algorithmic decisions should carry the same level of significance for all groups, such that for all students predicted to drop out, the same percentage of students in each group actually drop out. This is visualized in Figure 4 (right panel) which shows that 30 male and 35 female students are predicted to drop out, and 60% in each group actually drop out. The predictions therefore carry the same significance for male and female students. However, satisfying sufficiency may offer only a weak guarantee for fairness. To see this, suppose that the true dropout rate is 25% for males and 50% for females. An algorithmic system could naively predict all male students to be at-risk to achieve a precision of 25% and predict only a few female students to be at-risk to also achieve a precision of 25%. This algorithmic system satisfies sufficiency but it is clearly not fair. If additional resources were allocated to students predicted to be at-risk, then this system would have all of the male students receive this academic intervention, while withholding it from truly at-risk female students.

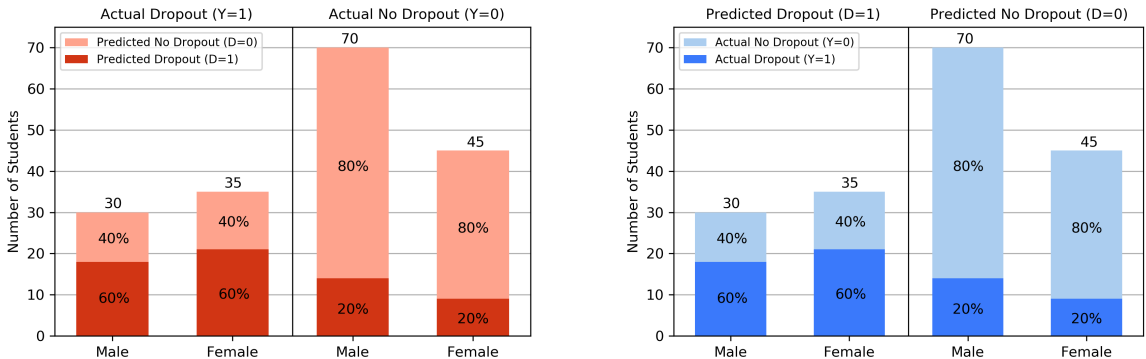

**Figure 4.** Illustration of dropout predictions that satisfy fairness as separation (left) and sufficiency (right) for a total of 100 male and 80 female students. Separation is satisfied with $Pr(D = 1|Y = 1, G = \text{male}) = Pr(D = 1|Y = 1, G = \text{female}) = 60\%$ and $Pr(D = 1|Y = 0, G = \text{male}) = Pr(D = 1|Y = 0, G = \text{female}) = 20\%$. Sufficiency is satisfied with $Pr(Y = 1|D = 1, G = \text{male}) = Pr(Y = 1|D = 1, G = \text{female}) = 60\%$.

*Similarity-based notions of fairness*

Statistical notions of fairness consider algorithmic decisions $D$, true outcomes $Y$, and group membership $G$, but they ignore all individual features of cases $X$. They are also known as group fairness measures, because they ignore individual differences. Whereas statistical notions of fairness require some kind of group-level parity, similarity-based notions of fairness require parity for pairs of similar individuals based on their observed features. We review two approaches to achieving similarity-based fairness: fairness through unawareness and individual fairness. Both approaches encode the belief that algorithmic decisions should not be influenced by any protected attributes that are irrelevant to the prediction task at hand.

**Fairness through unawareness** is an approach to achieve similarity-based fairness by omitting protected attributes from the feature set (Kusner et al., 2017). It avoids the appearance of disparate treatment because the algorithmic system does not take the protected attribute into account during model learning and action. However, this approach falls short of being blind to protected attributes, because a learned model can inadvertently reconstruct protected attributes from a number of seemingly unrelated features. For example, even if gender is removed from the feature set, a dropout prediction algorithm can still predict as if it had students' gender as a feature, because it has access to multiple features that are slightly correlated with gender. The consequences of fairness through unawareness can therefore be unexpected. Kleinberg and colleagues (2018) found that including race as a feature improves both overall accuracy and demographic parity of an algorithmic admissions system that predicts college success. By explicitly considering the applicant's race, the system's predictions of college success become more accurate and the fraction of black applicants who are predicted to succeed increases. Yu and colleagues (2020) found that predictions are more accurate if the feature set includes student demographic information in an algorithmic system that predicts student performance in college courses on the basis of learning management system data. In contrast, Yu and colleagues (2021) found that the inclusion of student demographics does not affect the overall performance and algorithmic fairness of a college dropout prediction model trained on registrar data for online and in-person college students. In light of these findings, it is unclear whether the inclusion of protected attributes adds predictive value to algorithmic systems, and the merits of fairness through unawareness may be mostly symbolic. Critics of the fairness through unawareness approach have noted its similarities to racist ideologies such as color-blindness (Bonilla-Silva, 2006; Burke, 2018).

**Individual fairness** goes one step further to address the issue of algorithms inadvertently reconstructing protected attributes from the feature set by quantifying the similarity between individuals directly (Dwork et al., 2012). This approach has domain experts construct a distance metric to capture the similarity between individuals for a given prediction task. Individual fairness requires that algorithmic decisions be similar for any pair of individuals that is close according to the task-specific distance metric. The similarity of algorithmic decisions is defined by the distance between the probability distributions over outcomes generated by a prediction model $f$. In the case of student dropout prediction, suppose that domain experts determine that academic preparedness for a course is the single best predictor of dropout and there exists an accurate measure of this student attribute. Students with similar levels of academic preparedness should then be classified similarly by the algorithmic system to satisfy individual fairness. In contrast, an unfair system, as depicted in Figure 5, predicts substantially different dropout probabilities for two similar students in terms of their academic preparedness, such that the discrepancy in predictions is not accounted for. A significant challenge with individual fairness is that it depends heavily on the choice of the distance metric, which itself can be subject to fairness issues. Moreover, treating similar individuals similarly may not produce outcomes that satisfy group-level fairness; it can for instance result in a rising tide scenario as shown in Figure 1. Specifically, if male and female students have the same tendency to drop out, then imposing individual fairness is a specific version of the group fairness notion of independence; but if the two groups differ in their likelihood of dropping out, then imposing individual fairness will violate independence.

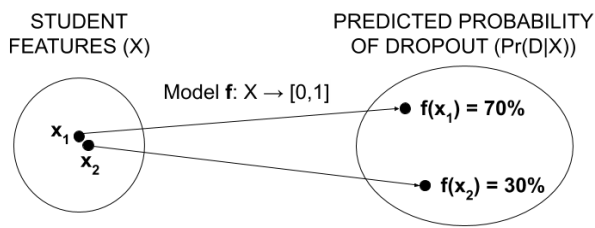

**Figure 5.** Illustration of dropout predictions that violate individual fairness. A prediction model $f$ maps two similar students $x_1$ and $x_2$ to different predicted probabilities of dropout, such that $x_1$ is predicted to be much more likely to drop out (70%) than $x_2$ (30%).

*Causal notions of fairness*

Statistical and similarity-based notions of fairness are based purely on observations of random variables, but algorithmic fairness can also be approached from a causal perspective. **Counterfactual fairness** is grounded in a causal approach to fairness. It encodes the belief that an algorithmic decision is fair if the prediction remains unchanged under the counterfactual scenario where the individual belongs to a different protected group (Kusner et al., 2017). Compared to individual fairness, which demands that pairs of similar individuals receive similar predictions, counterfactual fairness demands that any individual receives similar predictions regardless of their group membership. Evaluating how the individual's predictions would change for different group memberships requires causal inference. For example, a dropout prediction system that satisfies counterfactual fairness would predict the same dropout probability for a given student with their actual sex and with the opposite sex, keeping all other features constant. The challenge with this approach is making these causal inferences in the absence of a credible identification strategy, and especially if it requires extrapolating outside of the training set (i.e. no student in the training set with similar individual features is of the opposite sex) (Russell et al., 2017). Just as individual fairness hinges on the definition of a valid distance metric to measure similarity, counterfactual fairness hinges on the validity of a causal model for deriving causal quantities mostly from observational data (Wu et al., 2019).

This review of fairness definitions is not exhaustive but it covers fundamental notions of fairness that many of the definitions that have been proposed in the literature can be related to. An overview of the many proposed definitions of algorithmic fairness categorized by the fundamental notion they are related to is provided in Table 1. The various fairness definitions are either equivalent to or a relaxation of the statistical, similarity-based, and causal notions of fairness notions reviewed here.

**Table 1.** Definitions of algorithmic fairness categorized by general notions of fairness (Barocas et al., 2019; adapted from Verma & Rubin, 2018).

| Fairness notion | Related definitions of algorithmic fairness |
| --- | --- |
| Statistical fairness: Independence | Statistical parity/group fairness (Dwork et al., 2012); demographic parity (Feldman et al., 2015); conditional statistical parity (Corbett-Davies et al., 2017); Darlington criterion (4) (Darlington, 1971) |
| Statistical fairness: Separation | Equal opportunity/equalized odds (Hardt et al., 2016); ABROCA (Gardner et al., 2019); conditional procedure accuracy (Berk et al., 2018); avoiding disparate mistreatment (Zafar et al., 2017); balance for the positive/negative class (Kleinberg et al., 2016); predictive equality (Chouldechova, 2017); equalized correlations (Woodworth et al., 2017); Darlington criterion (3) (Darlington, 1971) |
| Statistical fairness: Sufficiency | Conditional use accuracy (Berk et al., 2018); predictive parity (Chouldechova, 2017); calibration within groups (Chouldechova, 2017); Darlington criterion (1), (2) (Darlington, 1971) |
| Similarity-based fairness | Fairness through unawareness (Kusner et al., 2017); individual fairness (Dwork et al., 2012) |
| Causal fairness | Counterfactual fairness (Kusner et al., 2017); no unresolved discrimination/no proxy discrimination (Kilbertus et al., 2017); fair inference (Nabi & Shpitser, 2018) |

*Choosing a Measure of Algorithmic Fairness*

The number and variety of available measures of algorithmic fairness raises an important question for researchers and practitioners alike: how does one decide on which measure to evaluate and monitor the fairness of a given algorithmic system? Answering this question requires careful consideration of how the algorithmic system is to be used (i.e. the action step): evaluating the fairness of a given system can require different measures depending on the use case, such as using dropout predictions for placing students on academic probation versus assigning them a personal tutor. To narrow down the options, one can choose to either ensure fairness at the group level using statistical notions or at the individual level using similarity-based and causal notions. Measures of individual-level fairness, such as similarity-based and causal fairness, provide more fine-grained information than group-level definitions, but they are generally harder to compute and implement. The extent to which they are feasible depends on the application context. For example, counterfactual fairness can be appropriate and feasible for an algorithmic system that delivers interventions to at-risk students with some degree of randomness to examine intervention efficacy, because the randomness offers an identification strategy for credible causal inference. Similarity-based fairness can also be feasible in applications where domain experts agree on a valid distance metric to quantify individual similarity. It is worth considering the possibility of applying individual-level fairness measures in education, where educational technologies facilitate randomization of course materials and where the availability of student data and frequent assessments facilitates the definition of robust distance metrics.

Group-level fairness is consistent with the ideal of allocating scarce educational resources to various groups of students in an equitable fashion. An advantage of group-level statistical notions of fairness is that they are easy to measure by computing conditional probabilities of different random variables. However, various statistical notions of fairness are inherently in conflict with each other, as formalized in the impossibility of fairness results (Chouldechova, 2017; Kleinberg et al., 2016), which have since been demonstrated in the contexts of automated English proficiency test scoring (Loukina et al., 2019) and university student at-risk prediction (H. Lee & Kizilcec, 2020). Since any two notions of statistical fairness cannot be satisfied simultaneously (except in unrealistic scenarios), one has to prioritize one notion of fairness over others. Which notion of statistical fairness should be prioritized in educational contexts? Independence advances the goal of allocating resources equally to all students regardless of their background or qualification. As we have seen, independence is more appropriate if there is an even playing field to begin with or if the action taken with the algorithmic system has negative consequences for individuals, but less appropriate otherwise. Separation and sufficiency, both considered merit-based fairness measures, advance the goal of allocating resources equally across *qualified* individuals independent of their group membership. Merit-based fairness measures are most appropriate in applications like student dropout prediction for allocating additional study resources, where the goal is to estimate students' true tendency to drop out as accurately as possible. Some algorithmic systems in education seek a middle ground between these two ideals of resource allocation and therefore need to satisfy a (weighted) combination of fairness criteria; for instance, an algorithmic admissions system is expected to rank the most qualified applicants highest while also maintaining socio-demographic diversity.

The choice of a fairness measure can also be informed by one's understanding of the relationship between students' true and observed attributes. Friedler and colleagues (2016) define two different worldviews that highlight the tension between group- and individual-level fairness: WYSIWYG (what you see is what you get) and WAE (we're all equal). The WYSIWYG worldview represents the belief that observed attributes (e.g., a student's GPA or standardized test score) are an accurate reflection of unobserved attributes (e.g., a student's merit or potential). In contrast, the WAE worldview represents the belief that there are no innate differences between groups of individuals in the unobserved attributes. In the example of college admissions, the WYSIWYG worldview assumes that observed attributes, such as GPA or standardized test scores, correlate well with unobserved attributes, such as merit or potential. This worldview encourages comparing the true aptitude of individual applicants using the observed student attributes, once any noise or structural bias in the observed attributes are appropriately accounted for. On the other hand, the WAE worldview assumes that different groups of applicants are all equal in their intrinsic abilities in the unobserved space, so any differences in the observed features are due to factors irrelevant to student ability (e.g., availability of resources for college preparation, or quality of neighborhood schools). Considering these two opposing worldviews makes the choice of fairness notions explicit, as only individual fairness notions are desirable under the WYSIWYG worldview, and only group fairness notions are desirable under the WAE worldview.

As different fairness notions all highlight different angles of fairness, it may not be sufficient to satisfy only one fairness definition in order to certify that an algorithmic system is non-discriminatory. Corbett-Davies and Goel (2018) point out a limitation of separation as a fairness criterion by showing that differential error rates across groups may just be a consequence of different underlying risk distributions, rather than a discriminatory algorithm. It can therefore be misleading to assess the fairness of a dropout prediction system solely based on separation without careful examination of the underlying distributions for each group, since a violation of separation may just be a natural consequence of male and female students having different probabilities of dropout. In some contexts, for example in automated essay scoring where an algorithmic system attempts to replicate a complex task performed by experts, it is possible to compare model predictions to reference values, such as the scores a human assigns to the same work. While human scores should not simply be treated as a gold standard because of biases that can enter in the scoring process, the comparison between human and algorithmic scores for various subgroups can offer an intuitive method to measure fairness in contexts such as TOEFL and GRE essay scoring (Bridgeman et al., 2009, 2012).

No single definition of fairness will be appropriate across all algorithmic systems in education. It is the responsibility of the researcher and practitioner to evaluate their specific situation and decide what criteria are most important to them (Makhlouf et al., 2020). It is also relatively easy to compute several group-level fairness measures to assess how an algorithmic system performs across various criteria. The effects of imposing different fairness criteria can be evaluated based on the long-term effects on affected populations (Liu et al., 2018), their trade-offs between social utilities (Corbett-Davies et al., 2017), their society-wide distributional effects (Hu & Chen, 2018), and changes in strategic behavior of affected individuals (Milli et al., 2019). This demonstrable need to carefully evaluate fairness criteria for each application scenario encourages active discussion about fairness priorities among a wide range of stakeholders.

**Advancing Algorithmic Fairness in Education**

Advances in algorithmic fairness in education rely on stakeholders raising questions about the presence and nature of disparate impacts of algorithmic systems in academic contexts. These questions apply to algorithmic systems that are currently in use, under consideration for procurement from external vendors, and new systems that analysts and researchers are planning to develop. Despite the recent enthusiasm about artificial intelligence, algorithmic systems are not silver bullets for solving problems in education (Reich, 2020). They are tools that in some cases impact many people, operate in opaque ways, and inadvertently cause harm in education as well as other domains (O'Neil, 2016). This is why it is critical for stakeholders to scrutinize the way that algorithmic systems are working in relation to how they expect them to work. We have seen how bias and discrimination can enter in every step of developing and deploying an algorithmic system, from measurement to model learning to action. We have also reviewed various notions of algorithmic fairness to measure how the system is actually working from different angles. We now offer recommendations for techniques that can be adopted at each step to improve algorithmic fairness in education. Fairness-enhancing techniques are an active area of research that has gained momentum over recent years (see Friedler et al., 2019 for an in-depth survey). The current best practice is to implement discrimination-aware unit tests at each step -- measurement, model learning, and action -- so that fairness issues can be identified and addressed in a timely and targeted manner (d'Alessandro et al., 2017). In the future, education technology providers may be expected to provide a fairness report to communicate benchmarked properties of their algorithmic systems to academic stakeholders (Mitchell et al., 2018).

To improve fairness in the measurement step, it is important to scrutinize the concrete prediction problem that the algorithmic system is set up to solve and remember that the underlying data is not neutral. Data reflect any existing biases and discriminatory behavior that exist in the real world. We reviewed several kinds of bias that can be present in the target measure as well as the feature set. Considering what is not included in the measured data can be just as important as what is included with respect to both missing information and sampling. Students in underrepresented groups may need to be oversampled to ensure that sufficient data is available in the model learning step to produce an accurate prediction for their group. Likewise, missing data may be more common among non-traditional students (e.g., standardized test scores, records from other learning technologies), which may inadvertently introduce bias in the model learning step. The collected data should therefore be carefully reviewed and corrected for potential bias before it is used to train prediction models. A number of techniques for de-biasing training data have been proposed (Calmon et al., 2017; Feldman et al., 2015; Kamiran & Calders, 2012). For example, Feldman and colleagues (2015) propose modifying individual features to standardize them across protected attributes to remove potentially discriminatory information encoded in the features. We encourage academic policymakers to interrogate the measurement step with questions about the definition of the prediction problem (e.g., "what is the system's definition of student success/failure?"), the data collection process (e.g., "what student data sources are used in the system?"), checks for bias in the training data (e.g., "how were students in the training dataset selected?"), and what de-biasing techniques were applied.

To improve fairness in the model learning step, developers of algorithmic systems in education can be explicit about which data pre-processing steps are necessary (Friedler et al., 2019), use model evaluation metrics that are sensitive to biases in prediction performance (Gardner et al., 2019), and add fairness constraints or regularizers to learning algorithms (Calders & Verwer, 2010; Kamishima et al., 2012; Zafar et al., 2017; Zemel et al., 2013). Friedler and colleagues (2019) find that different modifications to learning procedures result in a trade-off between model accuracy and fairness, and that the efficacy of a particular modification varies across different datasets. It is therefore important for developers of educational technology to carefully report and evaluate the choices made in model learning procedures, including pre-processing of student data as well as consideration of feature sets and learning algorithms. Model evaluation metrics like ABROCA (Gardner et al., 2019; Hutt, Gardner, et al., 2019) that account for group-specific model performance can reveal sources of discrimination for models that are trained to maximize overall accuracy. For example, differences in ABROCA values across student demographic groups can reveal a source of bias in a student predictive model that may otherwise be hidden if the model was only evaluated on standard metrics that consider the student sample as a whole. The model learning procedure can be modified by adding fairness constraints or regularizers to remove relationships with protected attributes that exist in the training data. For example, Dwork and colleagues (2012) suggest adding a fairness constraint to the optimization problem that maximizes accuracy and encodes the notion of individual fairness that similar individuals should receive similar predictions. We encourage academic policymakers to inquire about the use and consideration of fairness requirements in the model learning process and how its effects on students, instructors, and administrators are evaluated.

To improve fairness in the action step, it is important to remain vigilant about how the trained model is working for different stakeholders in education. If the model predictions do not achieve a desirable outcome for different groups, it is possible to improve the model in a post-learning step by modifying its predictions in a principled way (Hardt et al., 2016; Kamiran & Calders, 2012; Woodworth et al., 2017). For example, Hardt and colleagues (2016) suggest adjusting the resulting predictions of a trained model such that separation can be achieved post-hoc. Open-source toolkits such as Aequitas (Saleiro et al., 2018) and AI Fairness 360 (Bellamy et al., 2018) can be used in the action step to monitor a trained model for potential disparate impact by developers of algorithmic systems in education for internal audits and by education policymakers for external and periodic audits. Aequitas provides an interface for developers and policymakers to evaluate a prediction model based on several fairness metrics, along with a "fairness tree" to help them select a relevant metric for a given use case. AI Fairness 360 provides a more extensive toolkit that includes an interface for both detecting and mitigating unfairness of an algorithmic system. The design of these toolkits highlights the benefits of evaluating a variety of fairness measures in combination with a variety of bias-mitigation strategies. Last but not least, to improve fairness in the action step, algorithmic interfaces designed for students, instructors, and administrators can provide more information on the purpose and intended use of the algorithmic system and how it is working to promote trust in the system and avoid unintended types of usage that have adverse consequences (Conati et al., 2018; Doshi-Velez & Kim, 2017; René F. Kizilcec, 2016; Marcinkowski et al., 2020). Individual perceptions of algorithmic fairness, for example with respect to procedural and distributive justice, can diverge from objective measures of fairness based on how a system is presented to students in the context of peer assessment (Kizilcec 2016) or college admissions (Marcinkowski et al., 2020).

A vibrant new area of work on bias and discrimination in algorithms, which has become known as algorithmic fairness, is intersecting with new uses of algorithms and longstanding concerns about bias and discrimination in educational settings. The use of algorithmic systems in education is raising questions about its impact on students, instructors, institutions, and society as a whole. There is promising evidence that these technologies can benefit various stakeholders in different educational environments, but more work is required to fully understand the ways in which algorithmic systems that are in common use in education impact different groups or individuals differently. More critical analysis of AI in education that attends to its unintended and unforeseen negative consequences is necessary in light of the many recent examples of big data algorithms that reinforce pre-existing inequality (O'Neil, 2016). We anticipate many new investigations into algorithmic fairness in education in the coming years as the ideas and techniques reviewed in this chapter are gaining traction in our community.

*Risk Scores*. http://arxiv.org/abs/1609.05807

Kuhn, M., & Others. (2008). Building predictive models in R using the caret package. *Journal of Statistical Software*, *28*(5), 1–26.

Kusner, M. J., Loftus, J. R., Russell, C., & Silva, R. (2017). Counterfactual Fairness. In I. Guyon, U. V. Luxburg, S. Bengio, H. Wallach, R. Fergus, S. Vishwanathan, & R. Garnett (Eds.), *Advances in Neural Information Processing Systems 30* (pp. 4066–4076). Curran Associates, Inc.

Lee, H., & Kizilcec, R. F. (2020). Evaluation of Fairness Trade-offs in Predicting Student Success. *Fairness, Accountability, and Transparency in Educational Data Cyberspace Workshop*. Educational Data Mining (EDM).

Liu, L. T., Dean, S., Rolf, E., Simchowitz, M., & Hardt, M. (2018). Delayed Impact of Fair Machine Learning. In J. Dy & A. Krause (Eds.), *Proceedings of the 35th International Conference on Machine Learning* (Vol. 80, pp. 3150–3158). PMLR.

Li, X., Han, M., Cohen, G. L., & Markus, H. R. (2021). Passion matters but not equally everywhere: Predicting achievement from interest, enjoyment, and efficacy in 59 societies. *Proceedings of the National Academy of Sciences of the United States of America*, *118*(11). https://doi.org/10.1073/pnas.2016964118

Loukina, A., Madnani, N., & Zechner, K. (2019). The many dimensions of algorithmic fairness in educational applications. *Proceedings of the Fourteenth Workshop on Innovative Use of NLP for Building Educational Applications*, 1–10.

Luckin, R., & Cukurova, M. (2019). Designing educational technologies in the age of AI: A learning sciences‑driven approach. *British Journal of Educational Technology: Journal of the Council for Educational Technology*, *50*(6), 2824–2838.

Lum, K., & Isaac, W. (2016). To predict and serve? *Significance. Statistics Making Sense*, *13*(5), 14–19.

MacCarthy, M. (2019, December 6). *Fairness in algorithmic decision-making*. Brookings; Brookings.
26